\documentstyle[preprint,aps,psfig]{revtex}
\begin{document}
\bibliographystyle{unsrt}

\title{Source Size Scaling of Fragment Production in Projectile Breakup}

\author{L. Beaulieu$^{\rm{a}}$, D.R. Bowman$^{\rm{b}}$, D. Fox$^{\rm{b,}}$\footnote{Present 
address: Practical Political Consulting, East Lansing MI   48823, USA.}, 
S. Das Gupta$^{d}$, J. Pan$^{d}$, G.C. Ball$^{\rm{b}}$,
B. Djerroud$^{a,}$\footnote{Present address: NSRL, 
University of Rochester, N. Y., USA.}, D. Dor\'e$^{\rm{c,}}$\footnote{Present 
address: IPN Orsay, BP 91406 Orsay Cedex, France.}, A. 
Galindo-Uribarri$^{\rm{b}}$, D. Guinet$^{\rm{c}}$, E. Hagberg$^{\rm{b}}$, D. 
Horn$^{\rm{b}}$, R. Laforest$^{a,}$\footnote{Present address: AECL, Chalk
River Laboratories, Chalk River, Ontario, Canada, K0J 1P0}, Y.
Larochelle$^{\rm{a}}$, P. 
Lautesse$^{\rm{c}}$, M. Samri$^{\rm{a}}$, R. Roy$^{\rm{a}}$ and 
C. St-Pierre$^{\rm{a}}$}

\address{$^{a}$ Laboratoire de physique nucl\'eaire, D\'epartement de physique,
Universit\'e Laval,\\ Sainte-Foy, Qu\'ebec, Canada G1K 7P4.}
\address{$^{b}$ AECL, Chalk River Laboratories, Chalk River, Ontario, 
Canada K0J 1J0.}
\address{$^{c}$ Institut de Physique Nucl\'eaire de Lyon, 46 Bd du 11 Novembre 
1918, F-69622, Villeurbanne Cedex, France.}
\address{$^{d}$ Department of Physics, McGill University, 3600 University St., 
Montr\'eal, Qu\'ebec, H3A 2T8, Canada .}

\maketitle

\begin{abstract}
Fragment production has been studied as a function of the source mass and
excitation energy in peripheral collisions of $^{35}$Cl+$^{197}$Au at 
43 MeV/nucleon and $^{70}$Ge+$^{nat}$Ti at 35 MeV/nucleon. The results 
are compared to the Au+Au data at 600 MeV/nucleon obtained by the ALADIN
collaboration. A mass scaling, 
by $A_{source} \sim$ 35 to 190, strongly correlated to excitation energy per 
nucleon, is presented, suggesting a thermal fragment production
mechanism. Comparisons to a standard sequential decay model and the lattice-gas
model are made. Fragment emission from a hot, rotating source is unable to
reproduce the experimental source size scaling.
\end{abstract}

PACS number(s): 25.70.Pq, 25.70.Mn, 24.60.Ky

\vspace{0.5in}

Multiple emission of intermediate mass fragments (IMF), typically 
3$\leq$Z$\leq$20, also termed multifragmentation, is a well established decay 
mode in heavy-ion reactions, but theoretical descriptions are not 
straightforward, since many factors such as compression/expansion, temperature 
or instabilities~\cite{mor93a} may play a role. Attempts to isolate a thermal 
component from other effects in central collisions have been explored by 
Moretto {\em et al.}~\cite{mor95,pha95}. However, for such collisions, the 
properties of the emitters are not always well determined, as evidenced by 
the observation of binary collisions~\cite{lot92,que93,lec94,lar95} and 
neck emission~\cite{mon94,tok95,lec95}, which leave a very small cross section
for the formation of a single source~\cite{pet95,bea96}.

There are certain advantages in studying peripheral rather than central 
collisons. For example, compression effects can be neglected in a study of the
fast moving source formed in peripheral collisions. In particular, IMF 
production has been studied as a function of excitation energy in Xe, Au and 
U projectiles on gold targets at 600 MeV/nucleon by the ALADIN 
collaboration~\cite{tra95}. They showed that excitation-energy dependence of
the average IMF number for all three projectiles was the same
when scaled by the charge of the emitting source. This result goes beyond the 
target independence already found in Au projectile induced reactions at the 
same beam energy~\cite{hub92}, suggesting that the IMF production mechanism is 
independent of the emitter size. For a lighter projectile, such as $^{40}$Ca 
projectiles at 35 MeV/nucleon~\cite{des93,lle93}, the IMF emission was well 
reproduced by the sequential decay of a hot, rotating source~\cite{lle94} at 
variance with the ALADIN data~\cite{kre93}. This led the authors of 
ref.~\cite{des96} to consider the possibility of a size effect in the 
multifragmentation phenomena. However, the kinetic energies of 
IMF in the emitter frame from an argon projectile were not reproduced by the 
transition state formalism~\cite{jeo95}. Therefore even for light systems, a departure from standard sequential decay might be present. In this letter, we 
explore the effect of emitter size on IMF production for a much wider mass 
range than the ALADIN work~\cite{tra95}, namely
from A$\sim$ 35 to A$\sim$190 at excitation energies from 0.5 to 10 
MeV/nucleon. Comparisons are made to a standard sequential decay calculation
for a  hot, rotating source and to a lattice-gas model prediction.

The experiments were performed at the Chalk River TASCC facility with a
43 MeV/nucleon $^{35}$Cl beam on a 2.9 mg/cm$^2$ $^{197}$Au target and with 35 
MeV/nucleon $^{70}$Ge projectiles on a 2.1 mg/cm$^2$ $^{nat}$Ti target. Charged
particles were detected in the CRL-Laval forward array~\cite{pru90,lar94} 
consisting of 80 detectors mounted in 5 concentric rings around the beam
axis and covering the angular range from 6.8$^o$ to 46.8$^o$. 
The first three rings are made of fast-slow plastic detectors with 
charge resolution up to Z=20 and had thresholds
of 7.5, 12.5 and 16.2 MeV/nucleon for Z=1,6 and 10 respectively. The two
outer rings (24$^o$ to 46.8$^o$) are made of CsI(Tl) crystals with mass 
resolution for Z=1 and 2 and charge identification at Z=3. 
Ions with Z$\geq$4 are all attributed to Z=4. Thresholds were 2.5 MeV/nucleon 
for Z=1,2 particles. Finally, three Si-CsI(Tl) telescopes covered 
18\% of the solid angle between 3$^o$ to 5$^o$, with charge resolution from
Z=2 to 32 and typical thresholds of 2.5, 4.7 and 5.9 MeV/nucleon for Z=2,6 and 
10.

The fast-moving emitter in the $^{35}$Cl+Au peripheral reactions was selected 
by the iterative procedure described in section 3 of ref.~\cite{des93} for a 
system in the same mass range as ours. The data sample with total charge of 17 
consisted of 590000 events. In the case of the $^{70}$Ge+Ti reaction, 
separation of the moving source was more difficult, and each particle having a 
laboratory velocity, parallel to the beam, greater than or equal to 68\% of the 
beam velocity was attributed to the fast emitter. More than 
480000 events with total charge from 29 to 33 were selected. It was 
verified that the emission pattern was isotropic  in the emitter frame and 
that the kinetic energy spectra were well reproduced by a surface 
Maxwell-Boltzmann distribution~\cite{bea96b}.

The excitation energy was deduced, event by event, from the energy, angle and 
mass (from the charge) of each particle.
The number of neutrons was evaluated by a mass balance to allow correction to 
the excitation energy. Uncertainties in the excitation energy determination 
caused by contribution of pre-equilibrium nucleons are estimated to be 
a maximum of 7\% for 10 MeV/nucleon of excitation in the Cl data and up to 10\% 
for excitation energy higher than 8.5 MeV/nucleon in the Ge data. 
Pre-equilibrium emission does not change the conclusion of the present 
work~\cite{bea96b}.

The top panel of Fig. 1 shows the average number of IMF, $<N_{IMF}>$, as a 
function of excitation energy per nucleon for the Cl and Ge data. The ALADIN
Au data at 600 MeV/nucleon are also shown; the quantities of interest,
$<N_{IMF}>$, excitation energy and mass, were taken from ref.~\cite{tra95}. The 
definition of an IMF in the ALADIN data was 3$\leq Z_{IMF} \leq$30. For 
comparison with our results, the upper limit of $Z_{IMF}$ was scaled by the 
system size, 3$\leq Z_{IMF} \leq$30$\times (Z_{source}/79)$, 
giving 3$\leq Z_{IMF} \leq$6 for
Cl data and 3$\leq Z_{IMF} \leq$12 for Ge data. The size effect is clearly
seen as $<N_{IMF}>$ barely reaches unity for Cl, increases to about 2.2 for
Ge and goes beyond 4 in the ALADIN data. It should be pointed out that
the average mass of the emitting sources is almost constant in the cases of 
the Cl and Ge data because of our total charge requirement. For the Au+Au
reaction, the mass of the projectile spectator decreases from A$\sim$190 to 
A$\sim$50 as the excitation increases from 1.0 to 15 MeV/nucleon.

In order to remove the mass dependence from the data, average IMF numbers were
scaled by the source size for each bin of excitation energy. The striking
result of this operation is that all the curves coincide as
displayed in Fig. 1 (bottom panel). It must be noted that this universal 
scaling of fragment production applies to a wide range of masses, from
A$\sim$ 35 to A$\sim$190.  The beam energy dependence, from 35 MeV/nucleon to 
600 MeV/nucleon, is removed by using the excitation energy per nucleon.
The systems Xe+Au and U+Au at 600 MeV/nucleon exhibit similar behaviour 
\cite{tra95}. The new quantity, $<N_{IMF}>/A_0$, is strongly correlated with 
the excitation energy per nucleon, suggesting a thermal production mechanism
for the IMF.

Two different models have been used to explore the IMF production mechanism.
The first one is the lattice-gas model~\cite{pan95b,pan95}.  Given a 
freeze-out density and temperature the model can calculate the properties of 
the fragments.  This freeze-out density $\rho/\rho_0$ has been 
chosen to be as close as possible to 0.39, extracted from the analysis of 
Ar+Sc\cite{pan95b,pan95,tli93}.  The lattice dimension is 4$\times$5$\times$5 
for Cl, giving a 100 sites and a density ratio, $\rho/\rho_0$, of 0.35. The 
number of sites is 5$\times$6$\times$6 ($\rho/\rho_0$=0.39) for the Ge data.
The only free parameter left is the temperature. Secondly, IMF emission 
from standard sequential decay of a hot, rotating 
source has been simulated using GEMINI~\cite{gemini}. Based on the study of 
the $^{40}$Ca breakup at 35 MeV/nucleon~\cite{lle93}, a correlation between 
excitation energy and angular momentum has been used, up to the critical 
angular momentum values, which are 25$\hbar$ for Cl and 50$\hbar$ for Ge;
the correlation is determined such that the average IMF is best reproduced for each bin of excitation energy. Upon reaching the critical value, the angular momentum was kept constant as the excitation energy increased. Results from the two models were filtered for detector acceptance and thresholds, while all 
other quantities such the excitation energy and number of IMF were obtained following the same procedures used with the experimental data. 

The results are compared to the scaled IMF numbers in Fig. 2. In the upper
panel, the Cl simulations are displayed with the experimental data;
those for Ge are in the bottom panel. The general trend is very well 
reproduced by the lattice-gas model for both sets of data. 
In a standard sequential decay scenario, the simulations including effect due
to angular momentum reproduce the Cl data set well over the complete excitation range but deviate at energies above 5.5 MeV/nucleon for the Ge data.
Gemini predictions of $<N_{IMF}>/A_0$ reach their maximum around 0.030 
for the Cl data and $\sim$0.022 for the Ge data.
Within the framework of standard sequential decay, the predicted IMF production 
is strongly size dependent and therefore unable to reproduce the observed 
scaling. The same conclusion can be reached from comparison with the ALADIN 
data~\cite{hub92,kre93} where the calculated maximum of $<N_{IMF}>/A_0$ 
is even lower at 0.016~\cite{kre93}, showing that
GEMINI gives a decreasing value of the maximum of $<N_{IMF}>/A_0$ as the
mass $A_0$ increases, and fails to reproduce the experimental data at higher 
excitation energies. Therefore, GEMINI provides good agreement at low 
excitation energy, when it includes angular momentum. In such a model, the 
IMF production is dominated by angular momentum. For comparison, simulations 
with no angular momentum are shown; they underpredict the average IMF number. 
On the other hand, the use of the lattice-gas model might be unrealistic at 
low excitation energies since it incorporates features of 
prompt multifragmentation and phase instability; these features are 
essential to reproduce the high excitation energy data.

To insure that the observed scaling, in particular the good agreement of the 
lattice-gas model with the data, is not an artifact of the filtering process 
nor of the different selection methods used, unfiltered lattice-gas simulations 
were analysed. Within the theory, the temperature is related to excitation 
energy by the formula~\cite{pan95b,pan95}

\begin{eqnarray}
\frac{3}{2}T+\epsilon(N_{nn}^{max}-N_{nn}^T)/n=E^*_{QP}/A_0.
\end{eqnarray}
Here $E^*_{QP}/A_0$ is the excitation energy per nucleon.  
$N_{nn}^{max}$ and $N_{nn}^T$ are the number of nearest neighbour bonds
in the ground state and at temperature $T$ respectively.  The parameter
$\epsilon$ is related to the binding energy.  A value of $\epsilon $=3 $MeV$ 
is used in the present analysis.

The relation between temperature (divided by the critical temperature,
$T_c$=1.1275$\times \epsilon$ = 3.38 MeV) and excitation energy as calculated 
by Eq. 1 is shown in the top panel of Fig. 3. The results show a monotonic 
increase with $E^*_{QP}/A_0$. This is similar to experimental results recently 
obtained by the EOS collaboration for the Au+C reaction at 1 
GeV/nucleon~\cite{hau96,tin96}.  $<N_{IMF}>/A_0$ is also displayed (Fig. 3, 
bottom) for Cl and Ge and scaled the same way as the filtered simulations.
It shows no apparent size effects, in good agreement with the experimental
scaling.

In summary, a universal scaling has been presented for IMF production from
a wide range of source masses (35 to 190 nucleons) produced in reactions
with beam energies from 35 to 600 MeV/nucleon. This scaling is not reproduced
over the complete excitation energy range by the sequential decay of a hot, 
rotating source based on the transition-state formalism, suggesting
a possible change in the decay mechanism. The strong
correlation of $<N_{IMF}>/A_0$ with the excitation energy per nucleon and
the overall agreement of the lattice-gas model with the experimental data 
point to the thermal nature of IMF production. A thermal IMF production
mechanism is also consistent with recent results from the EOS 
collaboration, in which a continuous relation between temperature and 
excitation energy was found for the breakup of Au projectiles at
1 GeV/nucleon~\cite{hau96,tin96}. The extension of these results to 
very central collisions, where compression effects and instabilities are 
predicted, would put constraints on different models.

\acknowledgments
The authors would like to thank Dr. R.J. Charity for providing
his simulation code (GEMINI). This work was supported in part by the Natural
Science and Engineering Research Council of Canada (NSERC) and by
the Fonds pour la Formation de Chercheurs et l'Aide \`a la Recherche (FCAR, 
Qu\'ebec).

\pagebreak

\pagebreak

{\bf Figure captions}

\vspace{0.5in}

Fig. 1. Evolution of the average IMF number as a function of the excitation 
energy per nucleon (top panel). The open triangles represent the Cl data,
the black symbols the Ge data, and the crosses the ALADIN data taken
from ref. 13. In bottom panel,  the same results are scaled by the emitting
source size $A_0$. See text for details.

\vspace{0.5in}

Fig. 2. IMF number scaled by the source size as a function of the excitation 
energy per nucleon (same as the bottom panel of Fig. 1).
Top panel: Simulations for the Cl reaction with the lattice-gas model,
full line; GEMINI with correlation between excitation and angular momentum
up to 25$\hbar$, dashed line; and GEMINI without angular momentum, dotted line.
Bottom panel: Simulations for the Ge reaction; same symbols and line patterns
as above.

\vspace{0.5in}

Fig. 3. Unfiltered lattice-gas simulations of Cl and Ge. The relation
between the temperature, relative to the critical temperature($T_c$), and
the excitation energy obtained from Eq. 1 is shown in the top panel. IMF number 
scaled by the source size as a function of the excitation energy per nucleon
is presented in the bottom panel.

\newpage
\begin{figure}
\vspace{0.5in}
\psfig{figure=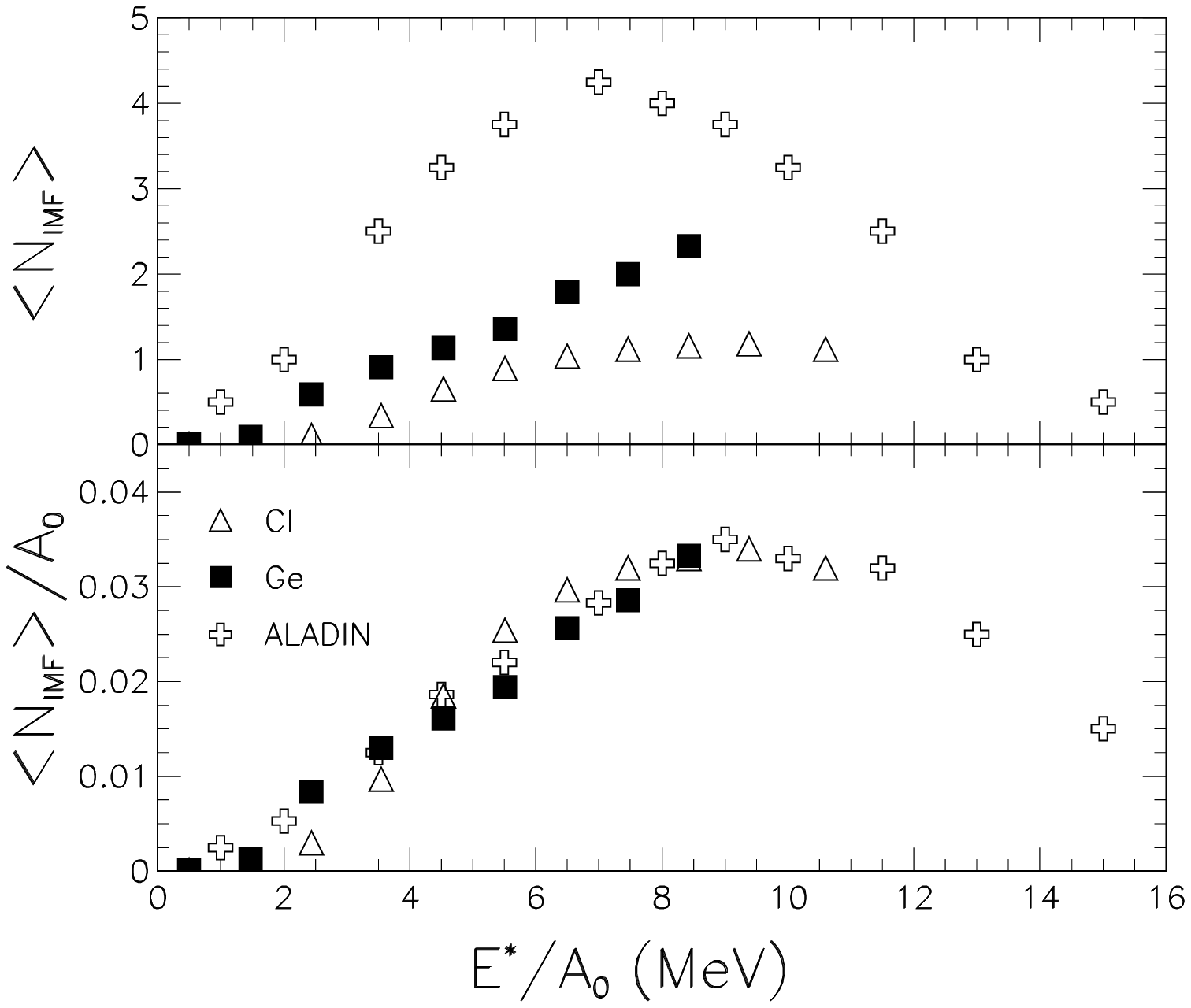,width=7in}
\vspace{0.5in}
\center{L. Beaulieu et al., FIG. 1}
\end{figure}

\newpage
\begin{figure}
\vspace{0.5in}
\psfig{figure=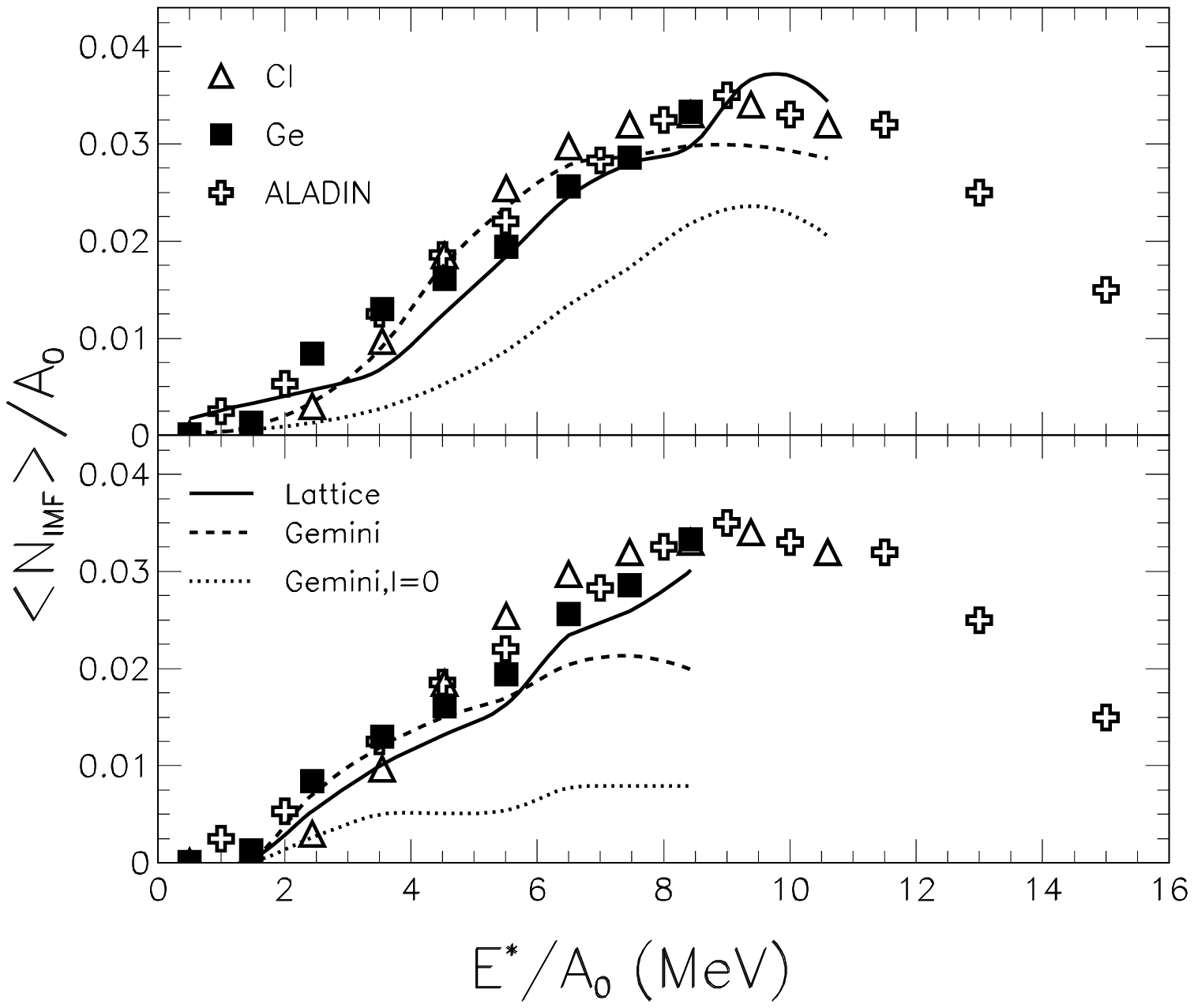,width=7in}
\vspace{0.5in}
\center{L. Beaulieu et al., FIG. 2}
\end{figure}

\newpage
\begin{figure}
\vspace{0.5in}
\psfig{figure=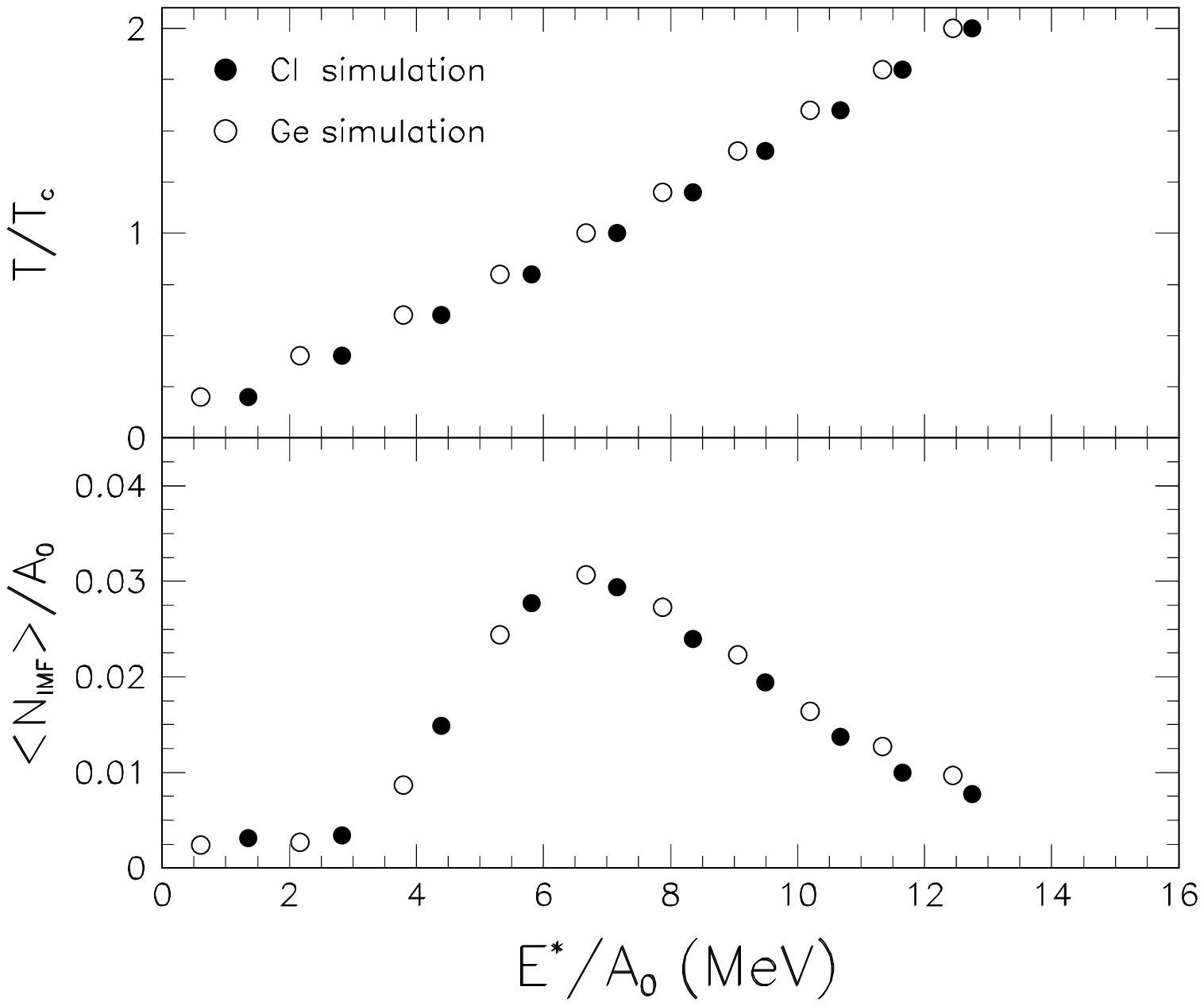,width=7in}
\vspace{0.5in}
\center{L. Beaulieu et al., FIG. 3}
\end{figure}

\end{document}